# Robust room temperature ferromagnetism in an itinerant van der Waals antiferromagnet

Longyu Lu[1], Qing Wang[2,3], Hengli Duan[4,5], Kejia Zhu[6], Tao Hu[1], Yupeng Ma[6], Shengchun Shen[7], Yuran Niu[8], Jiatu Liu[8], Jianlin Wang[9,10], Sandy Adhitia Ekahana[4], Jan Dreiser[4], Y. Soh[4], Wensheng Yan[5], Guopeng Wang[6,*], Yimin Xiong[6,11,*], Ning Hao[2,*], Yalin Lu[9,10,11] and Mingliang Tian[6,*]

[1]*School of Materials Science and Engineering, Anhui University, Hefei 230601, China*

[2]*Anhui Key Laboratory of Condensed Matter Physics at Extreme Conditions, High Magnetic Field Laboratory, HFIPS, Anhui, Chinese Academy of Sciences Hefei, 230031, China*

[3]*Science Island Branch of Graduate School, University of Science and Technology of China, Hefei, Anhui 230026, China*

[4]*Paul Scherrer Institut, 5232 Villigen PSI, Switzerland*

[5]*National Synchrotron Radiation Laboratory, University of Science and Technology of China, Hefei 230026, China*

[6]*School of Physics and Optoelectronics Engineering, Anhui University, Hefei 230601, China*

[7]*Department of Physics, University of Science and Technology of China, Hefei 230026, China*

[8]*MAX IV Laboratory, Lund University, Lund 22100, Sweden*

[9]*Hefei National Research Center for Physical Sciences at the Microscale, University of Science and Technology of China, Hefei 230026, China*

[10]*Anhui Laboratory of Advanced Photon Science and Technology, University of Science and Technology of China, Hefei 230026, China*

[11]*Hefei National Laboratory, Hefei 230028, China*

These authors contributed equally: Longyu Lu, Qing Wang, Hengli Duan, Kejia Zhu

* Corresponding authors.

*E-mail addresses*: 20239@ahu.edu.cn (G. Wang), yxiong@ahu.edu.cn (Y. Xiong), haon@hmfl.ac.cn (N. Hao), mltian@ahu.edu.cn (M. Tian).

**The coexistence of antiferromagnetic and ferromagnetic order at room temperature in single-phase van der Waals materials, particularly within the two-dimensional limit, has attracted significant research interest. Nonetheless, such materials are rare. In this work, we introduce an itinerant van der Waals antiferromagnet $(Fe_{0.56}Co_{0.44})_5GeTe_2$, where the ferromagnetic order of its exfoliated flakes remains discernible up to room temperature, extending down to the monolayer limit. A notable phenomenon observed is the evident odd-even layer-number effect at high temperature (e.g., $T$ = 150 K). Such behaviour can be**

**expounded by a linear-chain model. Of particular interest is the robust ferromagnetic order observed in even-layer flakes at low temperature (e.g., $T = 2$ K), which could potentially be attributed to spin-polarized defects. The intricate interplay among magnetic field strength, layer number, and temperature gives rise to a diverse array of phenomena, holding promise not only for new physics but also for practical applications**.

Layered van der Waals (vdW) antiferromagnets featuring intralayer ferromagnetic (FM) and interlayer antiferromagnetic (AFM) coupling have garnered substantial attention due to the manifold advantages, including negligible stray fields, stable and rapid spin texture dynamics, and emergent quantum phenomena[1–9]. The incorporation of FM order into the AFM host structure is crucial for achieving exchange coupling and potential spintronic applications[10,11]. In contrast to artificial systems constructed from assembled FM and AFM materials[12,13], the pursuit of a single-phase material with inherent coexisting FM and AFM behaviour, particularly one exhibiting intrinsic magnetic order, stands to amplify the potential functionalities and enhance convenience for commercial applications[14–16]. Notably, due to the odd-even layer-number effect in layered vdW antiferromagnets, the existence of robust FM in single-phase odd-layer materials becomes attainable, owing to the presence of an uncompensated layer[6,7,9]. However, a significant constraint faced by most known antiferromagnets lies in their propensity to exhibit insulating states, limiting their application potential in spintronics[2,5,7,17–19]. Clearly, itinerant vdW antiferromagnets emerge as a promising alternative due to the capacity to effectively mediate interlayer interactions and manifest robust spin-electron correlation[5]. Nevertheless, metallic antiferromagnets remain rare in the natural realm, especially those boasting a Néel temperature ($T_N$) surpassing room temperature.

Although predictions arising from the Mermin-Wagner theorem indicate that long-range magnetic order in two-dimensional (2D) materials should be suppressed at finite temperatures[20], substantial endeavors have been directed towards achieving robust FM ordering in ultrathin vdW magnets[1,21–29]. Recently, innovative approaches have emerged to counteract thermal fluctuations by employing magnetic anisotropy, achieved through mechanisms such as spin-orbital coupling or lattice distortion. These pioneering efforts have showcased intrinsic FM in various atomically thin vdW magnets, including $CrI_3$[1], $Cr_2Ge_2Te_6$[23], and $Fe_nGeTe_2$ ($n$ = 3, 4, 5)[5,24,30]. Additionally, the role of magnetic defects and impurities in determining magnetic and electronic properties cannot be overlooked. Through extrinsic defect engineering, the identification of localized magnetic polaron around a single atomic sulfur vacancy in $Co_3Sn_2S_2$ and the realization of unexpected magnetism in nominally non-magnetic $PtSe_2$ have been game-changing discoveries[31,32]. By synergizing magnetic anisotropy with magnetic defects, the prospects of achieving robust FM in layered AFM materials come into focus, irrespective of the odd-even layer-number effect. Recent studies have unveiled that the Curie temperature ($T_C$) of $Fe_5GeTe_2$ can be elevated to 320 K by partially substituting Fe sites with Co atoms. Further augmentation in Co doping triggers a complex magnetic phase transition to an A-type AFM ground state with $T_N$

~330 K[3,33,34]. Given the inherent existence of defects in as-grown materials, the potential for achieving potent FM in an AFM host is held by Co-doped $Fe_5GeTe_2$. Despite the promise, only a limited number of experimental inquiries have been undertaken.

In this study, we embark on a comprehensive exploration of the magnetic evolution of $(Fe_{0.56}Co_{0.44})_5GeTe_2$ (FCGT) across its layer spectrum, down to the monolayer extent. FCGT stands as an A-type itinerant vdW antiferromagnet with $T_N$ exceeding room temperature, as illustrated in Fig. 1a. Remarkably, monolayer form of FCGT exhibits a FM behaviour with an achievable $T_C$ of approximately 310 K—an impressive achievement when contrasted with that of numerous established exfoliated monolayer vdW magnets. Further intrigue arises from our investigations of the FCGT flakes. At $T$ = 150 K, these flakes exhibit a notable odd-even layer-number effect. This effect finds validation in both the discernible spin-flop transition and the thickness-dependent magnetization. An additional effect emerges in the form of even-layer flakes, which intriguingly showcase robust FM states comparable to their odd-layer counterparts at $T$ = 2 K. This phenomenon might well originate from spin-polarized defects. Through their interaction with conduction electrons, a surprising positive magnetoresistance (*MR*) is observed in both in-plane (IP) and out-of-plane (OOP) magnetic field orientations. The results presented here unveil a magnetic ground state within FCGT flakes that defies simplicity. These findings emphasize the potential for subsequent steps in the evolution of FM/AFM exchange coupling, thus paving the way for future spintronic applications.

## Results

**Room temperature FM in monolayer FCGT.**

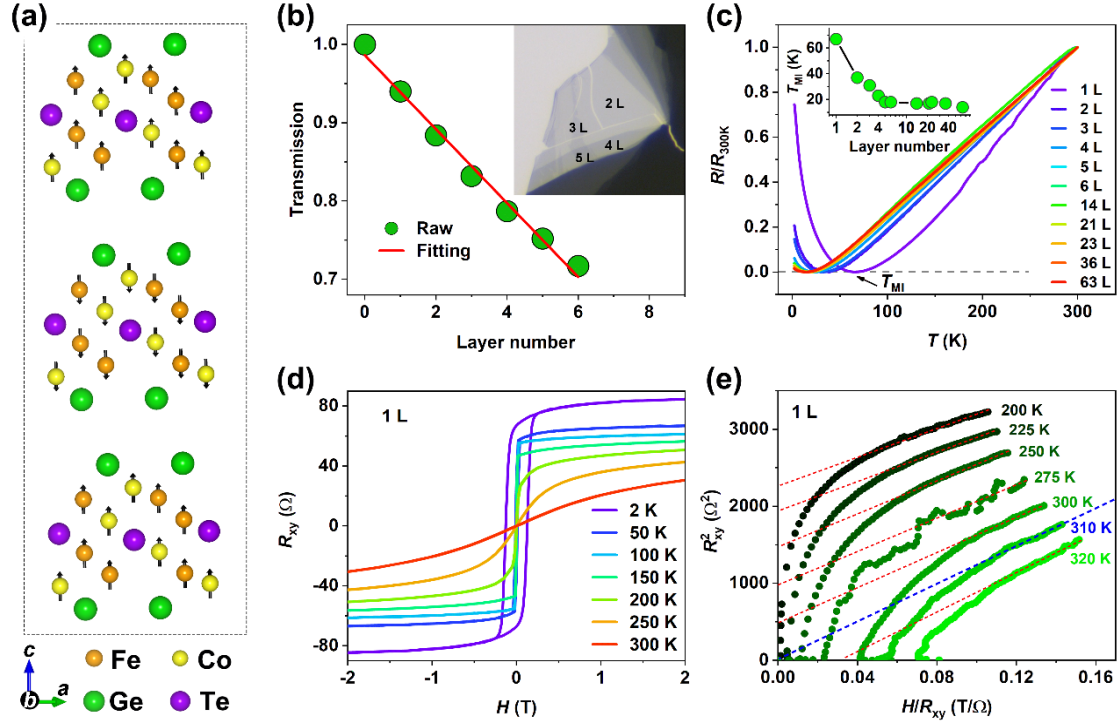

**Fig. 1 Monolayer FM of FCGT. a**, Crystal structure and magnetic order of 3 L FCGT. In this work, $Fe_3Co_2GeTe_2$ is utilized to simplify the calculations. **b**, Optical transmission as a function of the layer number. The red line is a fit to the data (green circles) using the Beer-Lambert law[24]. The inset shows a typical optical image of FCGT flakes exfoliated on a $SiO_2$/Si substrate. Regions with different layer numbers are marked out. **c**, Temperature-dependent longitudinal resistance of FCGT flakes with the layer number down to 1 L. All the samples exhibit metal-insulator transition. Resistances are normalized by the values at $T = 300$ K. The inset shows the metal-insulator transition temperature $T_{MI}$ as a function of the layer number. **d**, Anomalous Hall resistance $R_{xy}$ as a function of external magnetic field in a monolayer FCGT flake at various temperatures. **e**, Arrott plots of 1 L flake.

Before delving into the magnetic characteristics of FCGT flakes, we commence with the growth of high-quality bulk crystals using the chemical vapor transport method (Supplementary Fig. 1). The temperature-dependent magnetization ($M$-$T$) and $M$-$H$ analysis of the FCGT crystal affirm the presence of long-range AFM order at $T_N$ ~335 K (Supplementary Fig. 2a). The fitting of the inverse susceptibilities above $T_N$, the presence of non-linear $M$-$H$ curves at $T$ = 350 K, and the Hall measurements indicate an FM interaction comparable to or above $T_N$ (Supplementary Figs. 2 and 3). Element selective x-ray magnetic circular dichroism (XMCD) measurements on the bulk crystal confirm that the magnetic moments of Fe and Co atoms act in coordination with each other, with both contributing to the overall magnetization (Supplementary Fig. 4 and Supplementary Table S1). To ascertain the orientation of the magnetic easy axis of FCGT, we conduct IP $R_{xy}$ at varying temperatures and angle-dependent $R_{xy}$ measurements at $T$ = 2 K (Supplementary Fig. 5). The outcomes unambiguously unveil an OOP magnetic easy axis. This orientation proves to be pivotal in stabilizing long-range magnetic order in ultrathin flakes by introducing a substantial spin-wave excitation gap[1,23]. To delve into the realm of itinerant magnetism in the 2D limit, we fabricate devices using mechanically exfoliated FCGT flakes derived from bulk crystals, reduced to the monolayer scale. Determination of the exfoliated flakes layer count rests upon optical transmission and atomic force microscopy measurements (Fig. 1b and Supplementary Fig. 6). The normalized longitudinal resistance of FCGT flakes, varying in layer number, is exhibited in Fig. 1c. Remarkably, all these flakes retain metallic attributes even down to the monolayer extent, accompanied by a resistance upturn at low temperatures. This behaviour unveils a transition from a non-Fermi liquid state to the Kondo effect (Supplementary Fig. 7)[35]. In this context, the metal-insulator temperature ($T_{MI}$) is defined as the critical temperature at which resistance reaches a minimum. It is worth noting that $T_{MI}$ increases with decreasing layer number. In the 2D limit, electrons are more susceptible to being localized due to disorder than in 3D, rendering few-layer vdW materials more inclined towards an insulating behavior compared to their intrinsic metallic nature in bulk crystals[20,30,36–38]. Hence, the metallic behaviour exhibited by monolayer FCGT flake, with a low $T_{MI}$ ~67 K underscores the minimal structural disorder and high sample quality.

In Fig. 1d, the temperature-dependent $R_{xy}$ of the monolayer FCGT flake under OOP magnetic field is depicted. A nearly rectangular hysteresis loop featuring a characteristic negative butterfly-like MR (Supplementary Fig. 8) is evident, confirming the flake's FM nature with perpendicular magnetic anisotropy. As the temperature increases, the hysteresis loop gradually contracts and ultimately fades, indicating an FM-to-paramagnetic (PM) phase transition. Employing Arrott plot analysis precisely pinpoints the $T_C$ value, yielding $T_C$ (Arrott) of roughly 310 K (Fig. 1e). The Hall conductivity of the monolayer FCGT at low temperatures is approximately 100 S/cm, akin to that of pristine bulk crystals (Supplementary Fig. 9). Drawing on the well-established scaling theory of AHE, the Hall conductivity arises from intrinsic Berry curvature and extrinsic dirty metal scattering[39,40]. To validate the experimental findings, we embark on first-principles calculations for 1 L FCGT (Supplementary Fig. 10). A striking splitting in the spin-resolved band structure crossing the Fermi level signifies the presence of an itinerant FM state. The calculations yield a positive magnetic anisotropic energy (MAE, ~0.78 meV/u.c.), affirming the stabilization of long-range magnetic order along the OOP in 1 L FCGT.

**Strong FM in host AFM structure**

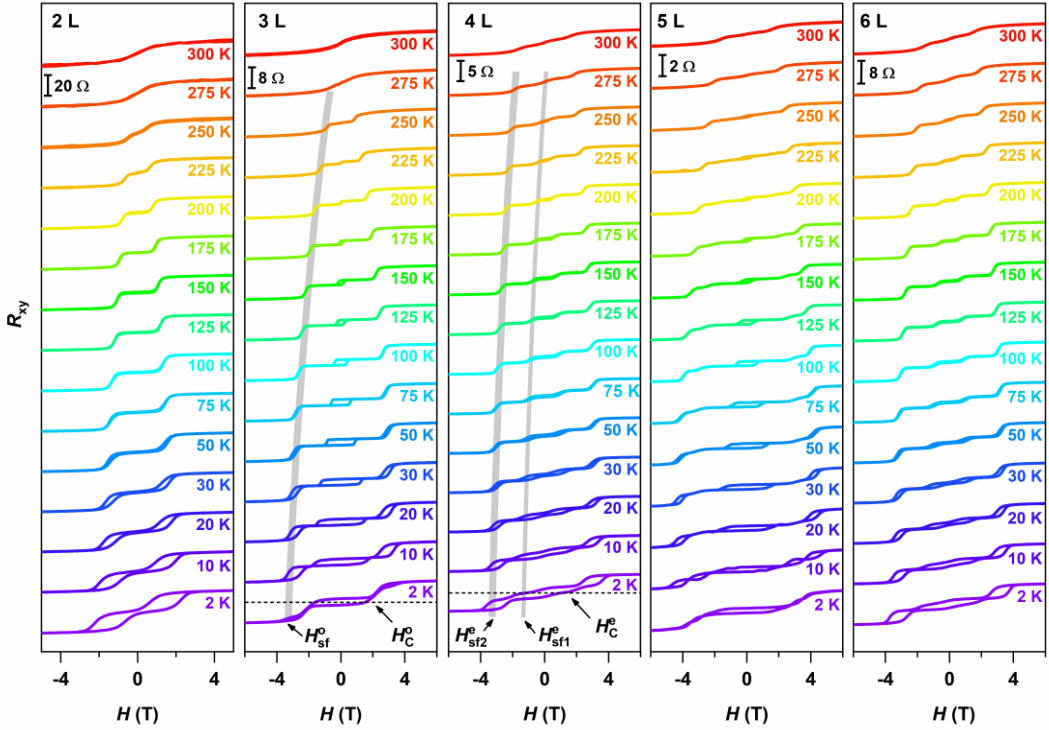

**Fig. 2 Spin-flop transition**. $R_{xy}$ as a function of magnetic field in FCGT flakes with layer number ranging from 2 to 6 L. The grey lines highlight the temperature dependence of the spin-flop transition in 3 and 4 L flakes. Dark scale bars denote the magnitude of the Hall signal for each device. The curves are offset for clarity. Here, $H_{\text{sf},1,2}^{\text{o/e}}$ denotes the spin-flop field and $H_{\text{C}}^{\text{o/e}}$ labels the coercive field.

We first investigate the layer-dependent AFM behaviour of FCGT flakes, focusing on those with layer number exceeding 1 L (Fig. 2 and Supplementary Fig. 11). Thanks to the distinctive A-type AFM configuration, a notable odd-even layer-number magnetic oscillation manifests at elevated temperatures (e.g., $T$ = 150 K), encompassing the layer number range from 2 to 21 L. Nonetheless, the oscillation becomes less discernible at lower temperature (e.g., $T$ = 2 K) due to the prominence of strong FM signals within the underlying AFM structure. Thus, to depict the layer-dependent AFM state, we employ the AHE at $T$ = 150 K. At this temperature, odd-layer flakes present a solitary hysteresis loop centered at $H$ = 0 T with a single spin-flop transition. This hysteresis loop confirms its FM nature (denoted as $FM_U$), originating from the presence of an uncompensated layer. The spin-flop field for odd layer flakes, $H_{sf}^o$, obtained from the derivative plot of $R_{xy}$-$H$, increases with the layer number (Fig. 3a and Supplementary Fig. 12). In contrast, even-layer flakes (with the exception of the 2 L flake) undergo a two-step spin-flop transition, which we label as sf1 and sf2. The spin-flop field $H_{sf1}^e$ for even layer flakes increases as the layer number increases, while $H_{sf2}^e$ slightly diminishes from 4 to 6 L before ascending with the layer number to 14 L. For the layer number surpassing 21 L, the magnetic behaviour closely resembles that of the bulk crystal, featuring a solitary spin-flop transition and a spin-flop field that remains insensitive to the layer number (illustrated as purple diamonds in Fig. 3a). This suggests a malleable interlayer exchange coupling by transitioning from a 3D bulk crystal to the 2D limit. As the temperature rises, the spin-flop transition gradually wanes. Consequently, a temperature-field diagram for both odd- and even-layer flakes is formulated (Supplementary Fig. 12), allowing us to infer that $T_N$ for layer number ≥ 4 L surpasses room temperature.

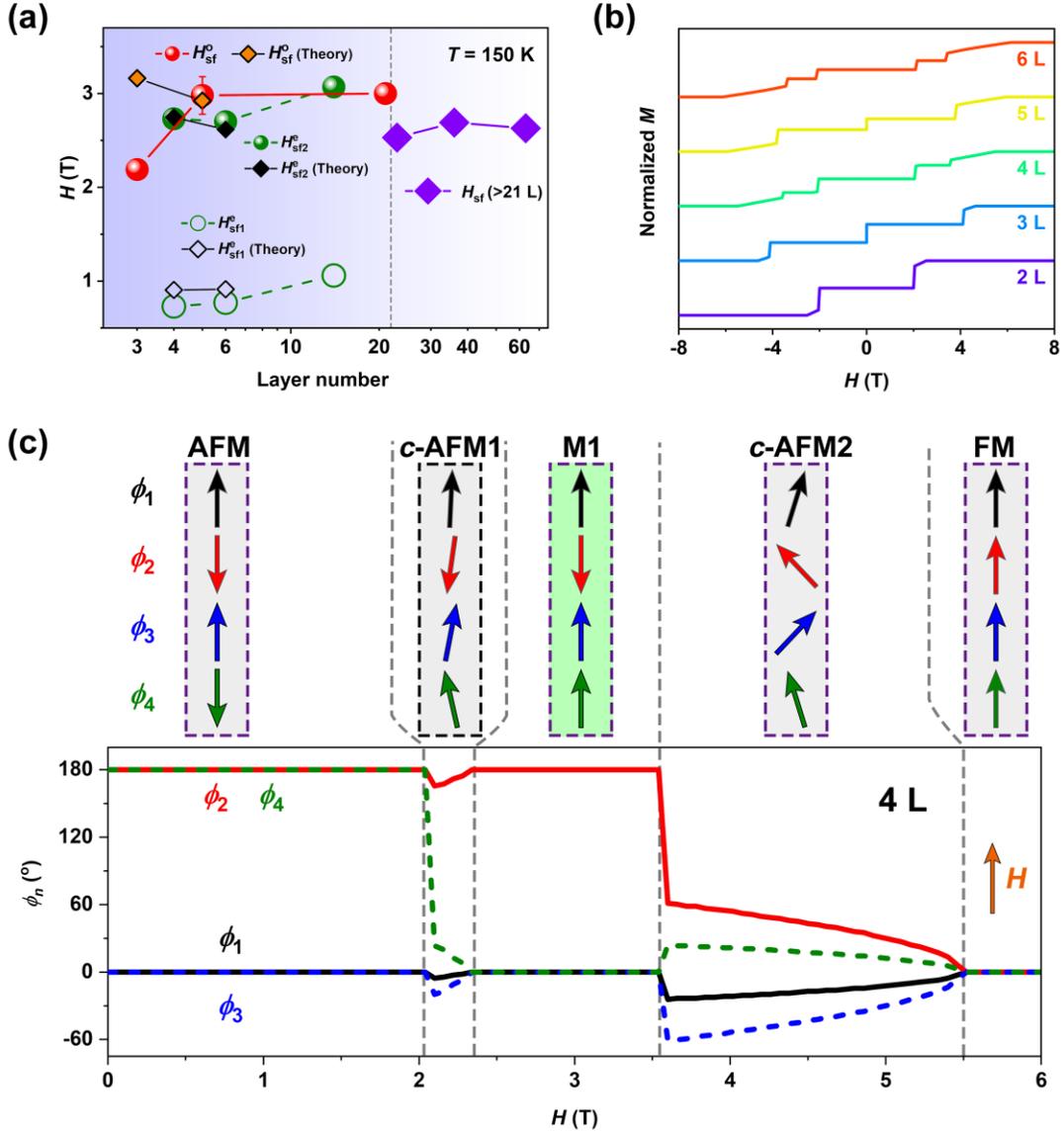

**Fig.3 Calculated spin-flop transition and magnetic state evolution. a**, Extracted experimental ($T$ = 150 K) and calculated spin-flop field ($T$ = 0 K) as a function of the layer number for odd- and even-layer flakes. For the layer number larger than 21 L, odd-even layer-number effect on the spin-flop transition becomes negligible. Here, the calculated $H_{sf}^o$ and $H_{sf2}^e$ are reduced by 1.3 times, while $H_{sf1}^e$ is reduced by 2.5 times. **b**, Calculated total magnetization from 2 to 6 L flakes. The magnetization of each flake is normalized by its saturation magnetization. The curves are offset for clarity. **c**, Calculated magnetic state evolution of the 4 L flake under an external magnetic field (pointing upward) using the AFM linear-chain model. The upper panel presents a schematic of spin reorientation from the AFM state to the FM state. The colored arrow orientations indicate the detailed magnetic moment alignments. The *c*-AFM1 and *c*-AFM2 represent the canted AFM1 and canted AFM2 states, respectively. The lower panel denotes the angle of "macrospin" in each layer. Each macrospin is marked with a different color. The angle $\phi_n$ is defined as 0° when the "macrospin" is pointing upward and 180° when pointing downward.

To understand the layer-dependent magnetism and the spin-flop transition, we employ the AFM linear-chain model to elucidate the magnetic state progression in FCGT. In odd-layer flakes, the magnetic evolution reveals the presence of a single spin-flop transition (Fig. 3a,b). Calculated values of the spin-flop field $H_{\text{sf}}^{\text{o}}$ exhibit a declining trend with increasing layer number. Although these calculated results differ from the experimental findings, this divergence might stem from the considerable influence of strong ferromagnetism within the ultrathin flakes on AFM interlayer coupling. Notably, even-layer flakes exhibiting a layer count ≥ 4 display a distinctive pattern with two spin-flop transitions. Intriguingly, the calculated spin-flop field variations align more closely with experimental observations (Fig. 3a,b). However, it should be pointed out that due to the simplified nature of our calculated structure (chemical formula $Fe_3Co_2GeTe_2$) and the fact that experimental results are acquired at $T$ = 150 K (while theoretical values are derived at 0 K), the absolute values of the calculated spin-flop fields deviate from experimental data. We extend our analysis of the magnetic state evolution in different layers of FCGT flakes, spanning from 2 to 6 L, as depicted in Fig. 3c and Supplementary Fig. 13. For a 2 L flake, as the magnetic field intensifies, the ground state changes from an AFM state ($\phi_1$ = 0°, $\phi_2$ = 180°) to a FM state ($\phi_1 = \phi_2 = 0°$), transitioning through a noncollinear state (canted AFM, *c*-AFM).

In odd-layer flakes with a layer count ≥ 3, the ground state adopts an uncompensated ferrimagnetic configuration. Preceding the attainment of saturation magnetic field featuring a FM state, the flake traverses the canted AFM state at $H_{sf}^{o}$. Notably, two spin-flop transitions are witnessed in flakes with a layer count ≥ 4. Following the completion of the initial spin-flop (c-AFM1), only the outermost layer ($\phi_4$ in Fig. 3c) with the "macrospin" in the "wrong" direction (with magnetic moment opposite to the external magnetic field) reverses its orientation to align with the external field, and the interlayer coupling with the adjacent layer ($\phi_3$) changes from AFM coupling to FM coupling. This is easy to understand since the outermost layer has only one adjacent layer coupled antiferromagnetically, so it takes a lower $H_{sf1}^{e}$ to align the moment to the external magnetic field compared to inner layers that have two adjacent layers coupled antiferromagnetically. This shift leads to the ground state assuming a ferrimagnetic M1 state instead of a FM state. In the M1 state, the number of $\phi_n = 0°$ exceeds that of $\phi_n = 180°$ by two. However, it's crucial to note that the M1 state emerges only when the ratio of $H_K$ (magnetic anisotropy) to $H_J$ (interlayer exchange interaction) surpasses a critical value (~0.33), as revealed in Supplementary Fig. 14. This critical point underscores the relatively robust magnetic anisotropy in comparison to interlayer exchange interaction in the context of vdW FCGT. The second flip-flop transition at $H_{sf2}^{e}$ is similar to the spin-flop transition for odd layer flakes.

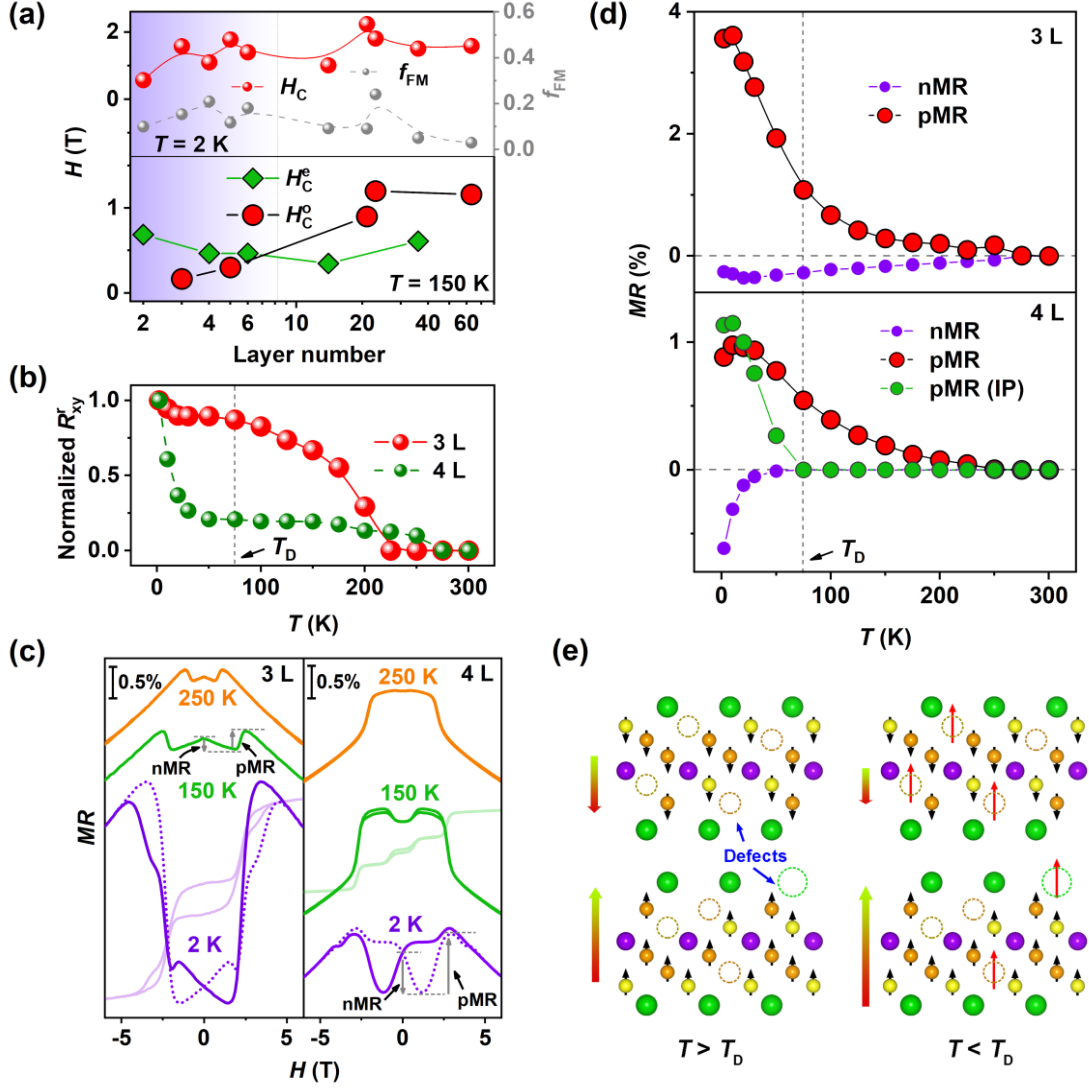

**Fig. 4 Strong FM in AFM flakes**. **a**, Top: Coercive field $H_C$ and fraction of FM component ($f_{FM}$) as a function of the layer number at $T = 2$ K. Bottom: $H_C$ versus layer number for odd- and even-layer flakes at $T = 150$ K. We use the remanent anomalous Hall resistance extracted from the normalized $R_{xy}$-$H$ data as $f_{FM}$. **b**, Remanent anomalous Hall resistance $R_{xy}^r$ as a function of temperature for 3 and 4 L flakes. The resistance is normalized by its value at $T = 2$ K. The grey dashed line marks the FM phase transition occurring at $T_D \sim 75$ K. **c**, Magnetic-field dependent magnetoresistance ($MR = 100 \times [R_{xx}(H) - R_{xx}(0)]/R_{xx}(0)$) at various temperatures for 3 and 4 L flakes. The magnetic field is perpendicular to the surface of the FCGT flake. At $T = 2$ K, the solid line (dotted line) represents a change in the magnetic field from 6 to -6 T (-6 to 6 T).

The grey arrows denote the negative (nMR) and positive MR (pMR), respectively. The corresponding $R_{xy}$-$H$ curves for 3 L (light purple line, $T$ = 2 K) and 4 L (light green line, $T$ = 150 K) flakes are plotted for comparison. The curves are offset for clarity. **d** Temperature-dependent extracted values of nMR and pMR components for 3 L (top panel) and 4 L flakes (bottom panel). The grey dashed line indicates the FM phase transition occurring at $T_D$ ~75 K. **e,** Schematic diagram of the spin texture for the 2 L flake at T > $T_D$ (left) and T < $T_D$ (right). The dashed circles denote the defects. The black and red arrows denote the Fe/Co and defect spins, respectively. The red-green arrows represent the total spin of a single layer of FCGT.

To comprehensively unveil the dimensional impact on the FM states, we undertake a meticulous analysis of the magnetic evolution. Interestingly, all measured flakes manifest hysteresis loops featuring discernible residual magnetic moments and substantial coercive fields $H_C$ at $T$ = 2 K (Fig. 2). This observation is unexpected for even-layer flakes given their inherent A-type AFM nature. The similarity in coercive field $H_C$ and $f_{FM}$ values between odd- and even-layer flakes provides evidence of a robust FM component in even-layer flakes (Fig. 4a, top panel). Furthermore, $H_C$ at $T$ = 2 K showcases a pronounced odd-even layer-number effect in the context of thin flakes (ranging from 2 to 6 L), where $H_C$ in odd-layer flakes slightly surpasses that in even-layer counterparts. This parity-dependent $H_C$ can be attributed to the additional Zeeman contribution from uncompensated layers[8]. Intriguingly, at $T$ = 150 K, an opposite odd-even layer-number effect emerges within the same thin flakes. $H_C$ in odd-layer flakes ($H_C^o$) becomes smaller than that in even-layer flakes ($H_C^e$), implying a magnetic transition between 2 and 150 K (Fig. 4a, bottom panel). Furthermore, $H_C^o$ at $T$ = 150 K exhibits a steady rise with increasing layer number, leveling off at 23 L. This behaviour arises from the requirement of a higher magnetic field to surpass the anisotropy energy in thicker odd-layer materials, allowing the Zeeman energy to exert dominance[6]. In contrast, $H_C^e$ at $T = 150$ K for even-layer flakes demonstrates a thickness-independent trend. This result can be ascribed to the net magnetization's origin from defects, which grows monotonically with thickness[6].

To further understand the intricate FM state, we delve into the temperature-dependent $R_{xy}^r$ behaviour, as showcased in Fig. 4b and Supplementary Fig. 15. Notably, odd-layer flakes (3, 5, and 21 L) exhibit an overall increase in magnetization as the temperature decreases, signifying a characteristic FM trait[28]. Conversely, even-layer flakes demonstrate a gradual ascent in $R_{xy}^r$ upon decreasing temperature, culminating in a sudden surge at $T_D \sim 75$ K. This abrupt increase suggests a magnetic phase transition and the emergence of a new FM state. This new FM state might also exist in odd-layer flakes and bulk configurations; this is corroborated by a minor upturn in $R_{xy}^r$-$T$ and a cusp in the $1/\chi$-$T$ curve at low temperatures (Supplementary Fig. 2a). The $R_{xy}^r$-$T$ behaviour of thicker odd-layer flakes (23 and 63 L) positions itself between the behaviour of odd- and even-layer samples. This is substantiated by the upturn below $T_D$, similar to even layer samples, possibly due to the uncompensated layer magnetization being a smaller fraction of the sample volume in the thicker samples. As the temperature increases, the $R_{xy}$-$H$ loops gradually fade, yielding either indistinct coercivity or zero-field remanence features at $T_C$ ($R_{xy}^r$) $\sim 300$ K for the layer number $\geq$ 5 L. This transition indicates a shift from a hard FM to a soft FM state characterized by multiple domains[24]. Arrott plots are employed to gauge $T_C$ (Arrott), which essentially marks the magnetic shift from the soft FM to the PM state. In 2 and 3 L flakes, the intercept at $T = 300$ K maintains a positive value, signifying a $T_C$ (Arrott) above room temperature (Supplementary Fig. 15c,d). Considering the universal thickness-dependent $T_C$ in 2D systems and the fact that the FM state exists above $T_N$ in 2 L, 3 L flakes, and FCGT bulk crystal, we reasonably infer that the $T_C$ (Arrott) of the remaining flakes (4 to 63 L) exceeds their $T_N$ with values above 300 K[1,23,24]. **This inference aligns with the observations of large FM domains on a micrometer scale via X-ray photoemission electron microscopy (PEEM) (Supplementary Fig. 16), further validating the room temperature FM behaviour.** Collectively, our analysis confirms the robust persistence of the strong FM state within the host AFM medium, even at room temperature. This discovery carries significant implications for both fundamental research endeavors and potential industrial applications.

**Mechanism behind the FM-AFM coexistence**

To gain insight into the origin of the FM states in FCGT flakes (≥ 2 L) and the interplay between spin states and conduction electrons, we conducted magnetoresistance (*MR*) measurements with the magnetic field along the OOP direction, as illustrated in Fig. 4c and Supplementary Figs. 17 and 18. A distinctive butterfly-shaped MR pattern is observed for all flakes at $T$ = 2 K, signifying the robust and potent FM state within the AFM host structure. Intriguingly, all FCGT flakes exhibit dominant positive *MR* (pMR) behaviour, paired with negative *MR* (nMR) near zero field, preceding the resistance drop at higher fields due to magnetization saturation. To understand the mechanism giving rise to the different types of *MR* behavior, we plot in Fig. 4c, the corresponding $R_{xy}$ (*H*) curve of the same flake. In Supplementary Fig. 17, we denote the spin-flop fields with a grey shade for flakes with layer number 3 and 4. The fields at which the *MR* switches from one type of behavior to another coincides with the spin-flop fields detected by tracking $R_{xy}$ (*H*), suggesting that the *MR* is correlated to the magnetic configuration. The resistance drop above the saturation field occurs at all temperatures and all sample thicknesses and can be explained by the reduced electron scattering in a better polarized magnetic medium.

As the temperature rises, both the pMR and the nMR linked to the FM states gradually recede. However, the temperature at which the nMR vanishes differs between odd- and even-layer flakes. From our previous work, the pMR could potentially stem from the magnetic easy axis along the IP direction[30]. To determine the orientation of the magnetic easy axis, we conduct angle-dependent Hall measurements on the 4 L flake at $T$ = 2 and 150 K (Supplementary Fig. 19a,b) and quantitatively analyze the magnetic anisotropy energy using the Stoner-Wohlfarth model (Supplementary Fig. 19c,d)[41]. The OOP magnetic easy axis is verified by the fitted positive first-order magnetic anisotropy energy $K_1$ of ~0.6 MJ/m$^3$ at $T$ = 2 K and ~0.59 MJ/m$^3$ at $T$ = 150 K, which effectively excludes the contribution of magnetic anisotropy to the pMR. By extracting pMR and nMR values, we construct plots to quantitatively analyze the *MR* evolution, as depicted in Fig. 4d and Supplementary Fig. 20. With increasing temperature, the pMR values for all flakes sharply decrease below $T_D$ ~75 K. This decrease is more pronounced in thicker flakes (36 and 63 L) and bulk crystals, where the pMR values nearly approach zero at $T_D$. Notably, the pMR in the IP direction completely vanishes at $T_D$ ~75 K (Supplementary Fig. 21). Specifically, angle-dependent AMR of the 4 L flake illustrates a gradual disappearance of pMR from the OOP to IP direction at $T$ = 150 K, while it persists in all directions at $T$ = 2 K (Supplementary Fig. 22). This suggests that some significant contribution to the pMR is correlated to the onset of FM below $T_D$. As for nMR, an evident odd-even layer-number effect emerges for the layer number < 21 L. This translates to the nMR persisting in odd-number layer flakes up to $T_C$ ($R_{xy}^r$), while it vanishes in even-number layer flakes at $T_D$. This observation aligns with the magnetic transition noted in Hall measurements, suggesting that it is associated with the zero remanent magnetization of the flakes. As for the pMR mechanism, we suspect that a different mechanism applies for T > $T_D$ vs T < $T_D$ based on the stark contrast in the anisotropy of pMR above and below $T_D$, displaying strong anisotropy above $T_D$ vs isotropic behavior below $T_D$. Due to the A-type AFM structure, Fe/Co atom spins are ferromagnetically coupled within a layer and antiferromagnetically coupled between the adjacent layers. This scenario leads to the suppression of spin fluctuations in layers with moments parallel to the magnetic field and possible augmentation in layers with moments opposite to the magnetic field under the OOP magnetic field, inducing pMR[42]. Conversely, when a magnetic field is applied in the IP direction perpendicular to the magnetic moments, MR approaches zero at low magnetic fields[42] (Supplementary Fig. 22b). This MR mechanism is strongly anisotropic, which may explain the anisotropic pMR above $T_D$. To understand the pMR below $T_D$, we consider, in addition to the mechanism described above, scattering associated to defects, which are introduced during the growth process, rather than magnetic impurities, as supported by EDS spectroscopy, STEM spectroscopy, and XPEEM measurements. As a result of defects, there is nonequivalent total spin at each layer, resulting in net magnetization (ferrimagnetism, $FM_F$) in even-layer flakes at T > $T_D$ (Fig. 4e, left panel). In odd-layer flakes, the total magnetization is a superposition of $FM_U$ and $FM_F$ components. Below $T_D$, we hypothesize that partial defects become polarized, adopting a FM configuration (denoted as $FM_D$). This polarization prompts a sharp surge in total magnetization, particularly in even-layer flakes (Fig. 4e, right panel). In dilute magnetic semiconductors, the introduction of magnetic ions prompts bound magnetic polarons through exchange interaction with localized carriers[7,43]. This notion is affirmed even in the Weyl semimetal $Co_3Sn_2S_2$, where the existence of localized spin-orbit polarons related to S-vacancies has been reported[31]. Drawing from previous findings and the fact that $T_D$ is significantly lower than $T_N$, suggests that defect moments could stem from magnetic polarons[7]. Angle-dependent Hall resistance measurements on an even-layer flake (4 L), showing substantial reduced remanent $R_{xy}$ as the tilt angle is increased at $T$ = 2 K, suggest that defect spins align along the OOP direction (Fig. 4e, right panel and Supplementary Fig. 19). The magnetic scattering of carriers from spin-polarized defects results in pMR observed in both OOP and IP directions, lowering the magnetic anisotropy of the pMR below $T_D$, which has contributions from the two different mechanisms described above. These local magnetic polarons and the associated pMR likely depend on the carrier density. The temperature dependence of carrier density follows a non-monotonic pattern (Supplementary Fig. 23), showcasing a dome-shaped behaviour with a peak near $T_D$. This indicates that spin carrier polarization (magnetization) and pMR increase as carrier density drops below $T_D$, consistent with previous observations in dilute magnetic semiconductors[43,44]. Our current understanding suggests that local magnetic polarons formation occurs below $T_D$. However, the specific defects contributing to magnetization and the origin of local magnetic polarons remain unclear, necessitating further research to elucidate these aspects.

**Discussion**

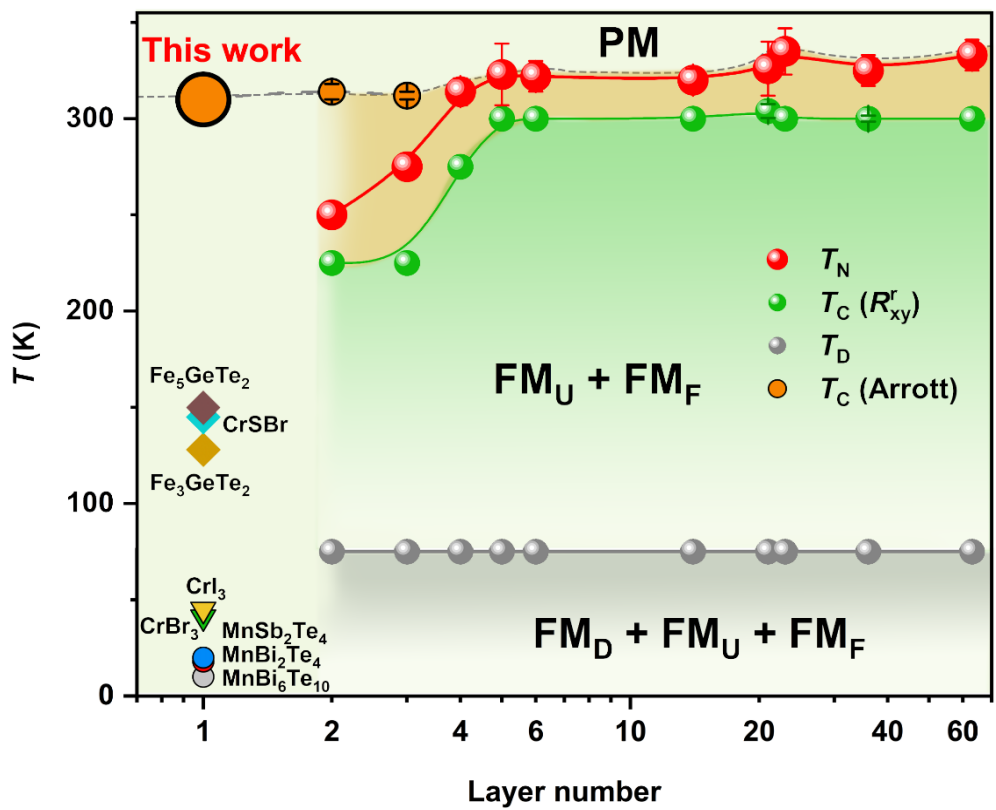

**Fig. 5 Temperature-layer number phase diagram of FCGT flakes.** Based on the preceding analysis, the $T_C$ (Arrott) of flakes spanning from 4 to 63 L should be larger than their $T_N$. We use the grey dashed line positioned slightly above $T_N$ to qualitatively serve as a boundary between the soft FM state and the PM state. The left panel shows a comparison of exfoliated single-layer vdW ferromagnets characterized by their Curie temperature $T_C$. The specific values of the data points are listed in Supplementary Table S2.

By combining the comprehension of FCGT flake's magnetic evolution, we delineate a temperature-layer number phase diagram, as depicted in Fig. 5. For layer number ≥ 2 L, the FM signal predominantly emanates from the uncompensated spin (odd-layer, $FM_U$) and spin-polarized defects ($FM_D$) at temperatures below $T_D$. It is important to note that not all defects are spin-polarized, implying the possible existence of $FM_F$ within this temperature range. Above $T_D$, the polarization of defects wanes, with $FM_F$ assuming dominance in even-layer flakes; in odd-layer flakes $FM_U$ is still present. As the temperature increases further, the hard FM state transitions into a soft FM state, eventually reaching the PM state with Curie temperature $T_C$ (Arrott) surpassing room temperature. In this context, $T_C$ (Arrott) is expected to marginally exceed $T_N$, where $T_N$ is deduced from a spin-flop field fitting (Supplementary Fig. 24). A noteworthy achievement is the intrinsic attainment of room temperature $T_C$ (Arrott) in the 1 L FCGT flake, very high in comparison to various exfoliated monolayer vdW ferromagnets. This intricate phase diagram underscores the distinct magnetic order in FCGT flakes, casting light on the potential for spintronic applications.

In summary, we trace the magnetic evolution of the room temperature itinerant antiferromagnet FCGT, unearthing an array of magnetic phase transitions extending down to the monolayer limit. Notably, a room temperature FM order accompanied by metallic traits is discerned in the monolayer flake. Within the framework of A-type AFM, a conspicuous odd-even layer-number effect surfaces at elevated temperatures (e.g., $T = 150$ K), evident through spin-flop transitions and oscillations in FM hysteresis loops. AFM linear-chain model provides a quantitative rationale for the experimentally observed odd-even layer-number effect in spin-flop transitions. The presence of spin-polarized defects fosters robust FM ordering in even-layer flakes at lower temperatures (e.g., $T = 2$ K). Remarkably, the magnitude of this FM state parallels that in odd-layer flakes, suggesting a novel avenue for robust FM achievement. Furthermore, the interplay of these spin-polarized defects with conduction electrons engenders intriguing pMR along both IP and OOP directions. These findings underscore the utility of FCGT flakes in unveiling a rich spectrum of exotic physics at the atomic scale, offering a distinct opportunity not only for exploring tunable magnetic attributes, but also for delving into the next-level FM/AFM exchange coupling and the emergence of novel quantum phenomena.

**Methods**

**Bulk synthesis and characterizations.** $(Fe_{0.56}Co_{0.44})_5GeTe_2$ single crystals were grown by the chemical vapor transport (CVT) method. The polycrystalline samples were firstly synthesized with Fe, Co, Ge, and Te powders in a molar ratio of 2.8 : 2.2 : 1 : 2. The polycrystalline samples were placed in a quartz tube with the addition of iodine as a transport agent. The quartz tube was vacuumed and sealed, and inserted into a double-temperature tube furnace for heating. Following heating at 750 ºC for 10 days and annealing in air, we obtained high-quality single crystals. The high crystallization of single crystals was confirmed by a room-temperature Rigaku 4-circle X-ray diffraction (XRD) diffractometer and their chemical composition was determined using energy dispersive spectroscopy (EDS) (Sigma 500, CarlZeiss). HAADF-STEM imaging and EDS mapping were conducted using a spherical aberration-corrected (Cs-corrected) 300-kV FEI Titan G2 microscope equipped with a Super-X detector. During the preparation of STEM samples, a focused ion beam (FEI Helios Nanolab 600i) was used. The thickness of the flakes was determined using atomic force microscopy (AFM, NX10, Park). Magnetization was measured using a SQUID magnetometer (MPMS, Quantum Design) under a magnetic field in IP and OOP configuration. The X-ray photoemission electron microscopy (XPEEM) and X-ray absorption fine structure (XAFS) measurements were employed at MAXPEEM and Balder beamlines (MAX IV Laboratory, Sweden), respectively. The Fe and Co $L$-edge XMCD measurements were performed at EPFL/PSI X-Treme beamline at the Swiss Light Source[45]. The magnetic field-dependent XMCD for both Co and Fe were obtained by measuring the difference between the TEY intensity at the $L_3$-edge peak and the pre-edge at each field for two different helicities.

**Device fabrication and transport measurements.** A Ti/Au (5/10 nm) standard Hall bar electrode was pre-fabricated on $SiO_2$/Si substrate using an ultraviolet lithography machine (MicroWriter ML3, DMO) and electron beam lithography (DSZ-500). The $(Fe_{0.56}Co_{0.44})_5GeTe_2$ flakes, which were mechanically stripped, were subsequently transferred to the Hall bar pattern using polydimethylsiloxane (PDMS). All the exfoliation and transfer processes were conducted inside an Ar-filled glove box ($H_2O$, $O_2$ < 0.1 ppm) to prevent any degradation. The devices were mounted into a custom-designed puck which was sealed with vacuum grease in the glove box to prevent oxidation before the transport measurements. Electrical transport properties were measured in standard four- or six-probe configurations using a commercial Quantum Design Physical Property Measurement System (PPMS-9T, Quantum Design) with the temperature mainly ranging from 300 to 2 K. To eliminate the offset caused by the misalignment of contacts in Hall resistance and longitudinal resistance, we utilized the data processing employed in previous work[24].

**First-principles calculations.** The first-principles calculations based on density functional theory (DFT) were performed with the projector augmented wave (PAW) method[46] and the generalized gradient approximation (GGA) developed by the Perdew,

Burke, and Ernzerhof functional (PBE)[47] using Vienna Ab initio Simulation Package (VASP)[48]. The wave functions of the valence electrons were calculated on a plane wave basis set with 520 eV cut-off energy. The cut-off energy was well convergent, which had been tested. The Brillouin-zone (BZ) sampling was executed by using a 15 × 15 × 1 Monkhorst-Pack $k$-point mesh for atomic geometry optimization and self-consistent calculations. A vacuum layer 20 Å thick was used to ensure decoupling between neighboring slabs, and the vdW corrections were included using the density functional theory including dispersion correction (DFT-D3) method[49] in our calculations of FCGT. All atoms were fully relaxed until the Hellmann-Feynman forces on each atom were smaller than 0.001 eV/ Å. The magnetic anisotropy energy (MAE) is defined as MAE = $E_{tot[100]} - E_{tot[001]}$, where $E_{tot[100]}$ and $E_{tot[001]}$ are the total energies of states with the spins parallel and perpendicular to the basal plane, respectively. Considering the real occupancy of the Fe1 positions, a 2 × 2 supercell was used to calculate the nearest interlayer exchange and magnetic anisotropic energy of the $Fe_3Co_2GeTe_2$ with 2–6 layers.

**Conflict of Interest**

The authors declare that they have no conflict of interest.

**Acknowledgments**

G.W. acknowledges financial support from National Natural Science Foundation of China (Grant No. 12104007). J.W. acknowledges financial support from National

Science Foundation of China (Grant No. 12004366) and National Key R&D Program of China and the Ministry of Science and Technology of China (Grant No. 2019YFA0405600). Y.L. acknowledges financial support from National Science Foundation of China (Grant No. 51627901). Y.X. acknowledges financial support from the Innovation Program for Quantum Science and Technology (Grant No. 2021ZD0302802). N.H. acknowledges financial support from National Key R&D Program of China (Grants No. 2022YFA1403200 and No. 2017YFA0303201), the National Natural Science Foundation of China (Grants No. 92265104, No. 12022413, and No. 11674331), the "Strategic Priority Research Program (B)" of the Chinese Academy of Sciences (Grant No. XDB33030100), and the Major Basic Program of Natural Science Foundation of Shandong Province (Grant No. ZR2021ZD01). M.T. acknowledges financial support from National Key R &D Program of the MOST of China (Grant No. 2022YFA1602603) and National Natural Science Foundation of China (Grants No. U19A2093). H.D. acknowledges financial support from National Natural Science Foundation of China (Grant No. 12105286). We acknowledge MAX IV Laboratory for time on Beamline MAXPEEM under Proposal [20221569] and Beamline Balder under Proposal [20220660]. Research conducted at MAX IV, a Swedish national user facility, is supported by the Swedish Research council under contract 2018-07152, the Swedish Governmental Agency for Innovation Systems under contract 2018-04969, and Formas under contract 2019-02496. This project also has received funding from the European Union's Horizon 2020 research and innovation programme under the Marie Skłodowska-Curie grant agreement No 884104 (PSI-FELLOW-III-3i). SAE acknowledges European Research Council HERO Synergy grant SYG-18 810451.

**Author contributions**

Guopeng Wang, Yimin Xiong, and Mingliang Tian conceived and designed the research project. Guopeng Wang and Longyu Lu performed the device fabrications and transport measurements. Kejia Zhu synthesized and characterized the bulk crystal, assisted by Tao Hu, Jianlin Wang, and Yalin Lu. Hengli Duan, Wensheng Yan, Yuran Niu, Jiatu Liu, Sandy Adhitia Ekahana, and Jan Dreiser performed the XPEEM, XMCD, and XAFS measurements as suggested by Y. Soh (for measurements at the SLS). Qing Wang and Ning Hao provided the theoretical calculation and the analysis. Guopeng Wang, Longyu Lu, Ning Hao, Y. Soh, and Mingliang Tian wrote the manuscript with input from all the authors. All authors discussed the results and commented on the paper.